# PACKAGING OF RF MEMS SWITCHING FUNCTIONS ON ALUMINA SUBSTRATE

Mohamad EL KHATIB, Arnaud POTHIER and Pierre BLONDY

XLIM Research Institute, 123 AV. A. THOMAS 87060 LIMOGES – FRANCE

## ABSTRACT

The expending development of wireless communication requires strong demands for components with improved capabilities. RF MEMS devices offer a variable alternative to conventional communication components because they consume less DC power, have lower losses, higher linearity and higher Q factor. However, the commercialization of RF MEMS devices is hindered by technological issues such as their packaging, which is the most critical problem to be solved. Compared with conventional IC packaging technologies, packaging of RF MEMS is for more challenging, because it implies both electrical and mechanical aspects. The mechanical nature of RF MEMS requires low temperature, hermetic wafer level packaging [1,2]. An interesting approach consist in encapsulating the component or a group of components on ceramic alumina substrate using simple techniques with limited technological stages. Our approach for the packaging of MEMS based switching networks consist in the encapsulation of the whole switching function instead of each MEMS switch alone, which simplify the design in terms of size and costs.

## 1. BASIC MEMS SWITCHING ELEMENTS

MEMS devices can easily be integrated on a large variety of substrates (silicon, AsGa, quartz…). In this work, we present designs on alumina substrate. The switching devices that we are presenting in this work are designed on alumina substrates, since many microwave applications are using this type of substrates and they can be easily post-processed, once the MEMS fabrication sequence was finished. Fig.1 shows a SEM and a schematic view of the basic MEMS switch integrated in the switching networks presented in this paper. It is a DC contact switch with a cantilever structure made in gold (150μm in length, 120μm in with and a thickness of 3.5μm). this types of switches shows good performance on a relatively wide frequency range (from DC to 30 GHz) and is much less prone to self-actuation compared to standard capacitive switches [3].

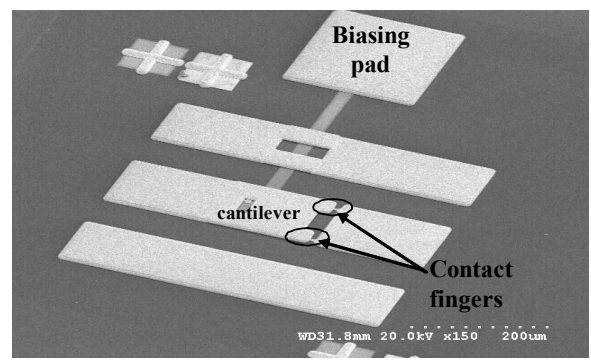
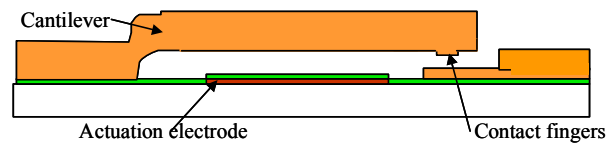

**Fig. 1: Schematics and SEM views of basic DC contact MEMS switch.**

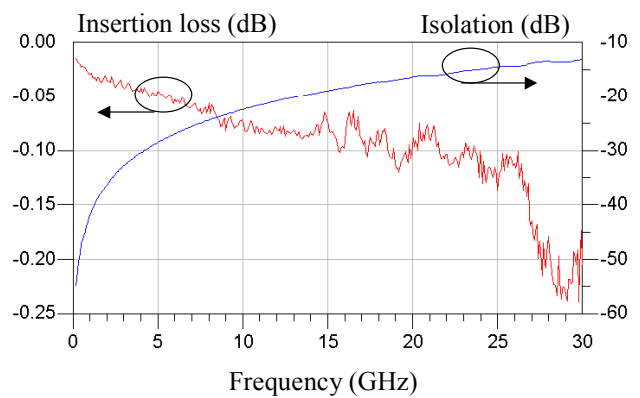

**Fig. 2: Measured isolation (OFF state) and insertion loss (ON state) of a DC contact switch implemented in CPW technology on alumina substrate.**

This device can be implemented as well in coplanar and in microstrip technology. However, the biasing network

Mohamad EL KHATIB, Arnaud POTHIER and Pierre BLONDY
PACKAGING OF RF MEMS SWITCHING FUNCTIONS ON ALUMINA SUBSTRATE

implementation is generally simplified with microstrip approach. The device is electrostatically actuated by a resistive electrode that prevents RF leakages in the biasing network, and is covered with a protective alumina dielectric layer. In order to have an adequate isolation in the OFF state, the metal to metal contact area during actuation was reduced to two small fingers, as shown on Fig 1. 0.3 µm thick dimples have been also added under the cantilever contact areas to improve long term contact reproducibility and to reduce the associated contact resistance. Fig 2 shows the typical performance obtained with DC contact MEMS switches fabricated on alumina substrate. Isolation in the OFF state is better than 38 dB at 2 GHz and 16 dB at 20 GHz. In the ON state, insertion loss is better than 0.3 dB from DC to 30 GHz, with an applied tension of 45-50 V.

## 3. PACKAGING APPROACH

MEMS devices are intented to be packaged in a micro-cavity fabricated by micromaching of a low cost glass substrate, which is directly bonded on the alumina wafer using gold-to-gold thermo-compression technique, avoiding any biasing or transmission line feedthrough underneath the sealing ring. The glass cap allows MEMS components to be hermetically encapsulated at the wafer scale in a controlled atmosphere. The RF coplanar access will be patterned on the lower alumina substrate side and will be connected to the MEMS active function built on the other subtrate side, through laser drilled full metallized vias holes, These coplanar access will facilitate the packaged component report using flip-chip technics . the via hole process in alumina substrate is now a mature and a very well controlled technology: several companies provide standard alumina substrates already laser drilled with metallized vias.

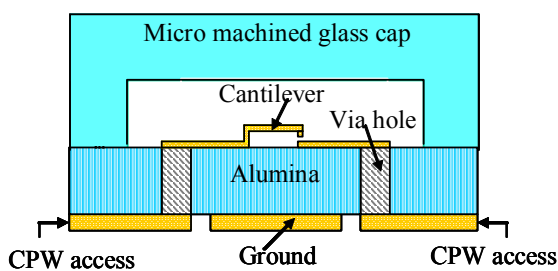

**Fig. 3 Side view of the RF MEMS package.**

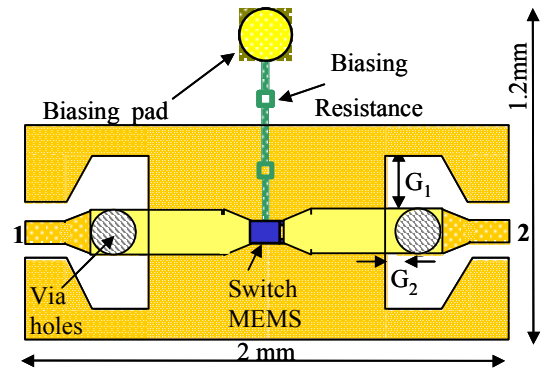

**Fig. 4: Top view of the packaged SPST component.**

This packaging approach remains compatible with additional circuit fabrication on the same wafer (filters, phase shifters, antennas…) and does not necessarily require individual cap cutting or etching. Fig 3 and 4 present side and top views of SPST(Single Port Single Throw) based on a single MEMS switch encapsulated as we described above. The micro-package RF performance depends mainly on the RF transitions used through the substrate. In fact, the RF transition matching could be easily adjusted by optmising the two gap values G1 and G2 between via holes and the CPW ground plane. In fact vias holes present an inductive influence as the frequency increases; the effect of this parasitic inductance on the package performance could be compensated ajusting the capacitance between vias holes and ground plane. The input impedance of the CPW access introduce also an important influence that has been optimized to reach the performances presented on Fig 5 in the SPST case.

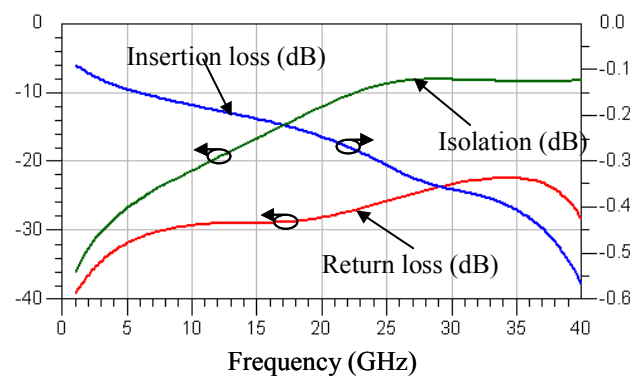

**Fig. 5: SPST simulated RF performance.**

In the ON state (transmission state), the SPST matching is better than 35 dB at 2 GHz and 22 dB at 35 GHz with a loss level lower than of 0.55 dB from DC to 40 GHz. In the OFF state, the MEMS switch is open with a 1.8µm gap between the movable cantilever and the contact fingers. The resulting isolation is better than of 20 dB up to 15 GHz and increase until 7 dB up to 40 GHz. The isolation level is influenced mainly by the switch geometry.

### 4. MULTIPLE OUTPUTS SWITCHING FUNCTIONS TOPOLOGY

In the following we present the design of several basic switching functions packaged using the same principle applied for the previous device. Fig.6 shows a SPDT device (Single Port Double Throw) that uses two DC contact switches encapsulated in the same package.

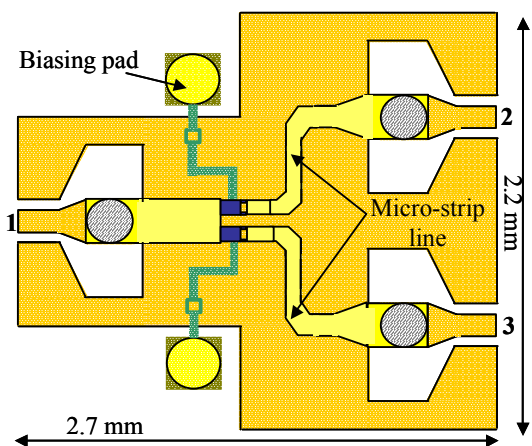

**Fig. 6 Top view of the designed SPDT with its biasing network.**

In the ON state the SPDT present a matching of 40 dB at 2 GHz, and better than 28 dB up to 30 GHz with a transmission loss level of 0.1 dB at 2 GHz lower than 0.9 dB up to 30 GHz. In the OFF state, the SPDT isolation between the input and output access, stays always below 17 dB from DC to 40 GHz and stays bellow 16 dB between the two output access.

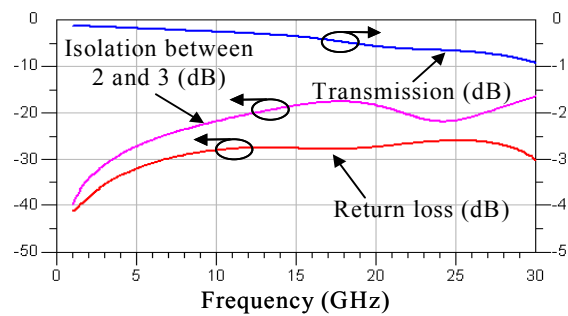

**Fig. 7: SPDT simulated RF performance, transmitting from 1 to 2.**

A Single Port Three Throw (SP3T) has been also designed and is presented on Fig 8. This component allow to direct the signal from the input access to any of the three outputs by actuating the corresponding switch (i.e. switch 2 permits a signal transmission from access 1 to 3), while the others switches are left in their OFF state.

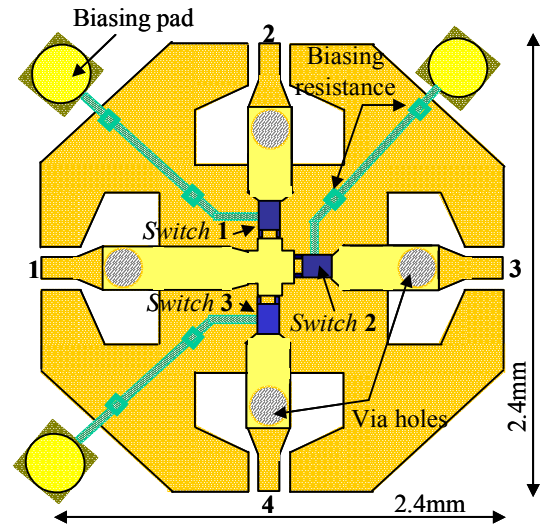

**Fig.8 Top view of the designed SP3T with its biasing network.**

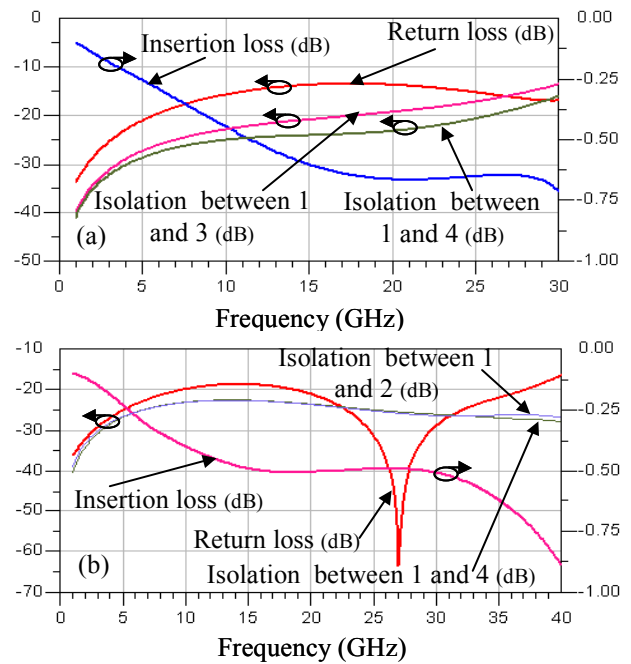

**Fg.9: Simulated RF performance of the SP3T topology (a) switch 1 activated and (b) switch 2 activated**

The simulation results for this device are shown on Fig 9. When the switch 1 is actuated, the transmission loss level stays below 0.1 dB until 2 GHz and 0.6 dB up to 30 GHz (fig.9.a). whereas when switches 2 or 3 are actuated, the simulated insertion loss is evaluated to be below 0.1dB until 2GHz and 0.7dB up to 30GHz (fig.9.b). This small degradation of the device performance could be explained by a small mismatching at the cross junction. The predicted RF performance in the OFF state seems to be correct until 20 GHz with isolation level better than 15 dB between the regarding RF access. Based on previous results for the basic devices, more complex component can be designed. The SP4T (Single Port Four Throw) example presented on Fig 10, is obtained by combining 3 SPDT devices. The global geometry has been optimized in order to minimize the size of the final structure and improve its performance as much as possible.

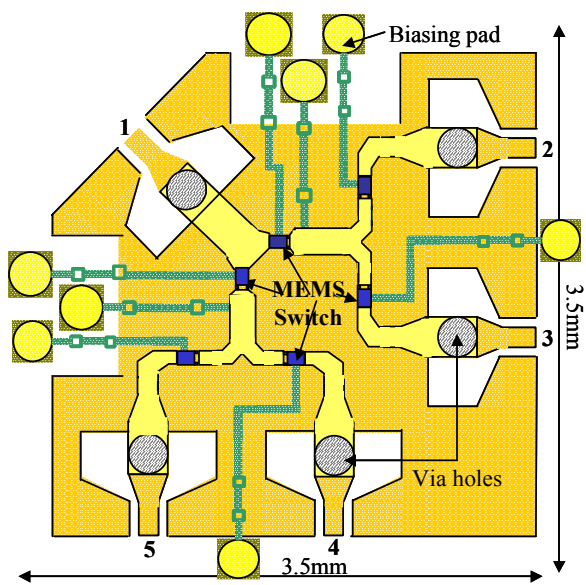

**Fig. 10: Top view of the optimized SP4T topology.**

In this specific topology, two DC contact switches need to be actuated at the same time to allow the signal transmission to any of the four-output access. This series combination will result in an isolation level improvement as shown on Fig. 11. On the other hand, the global transmission loss on each trajectory will increase due to the addition of the actuated individual MEMS switch loss. Consequently, the SP4T have been specifically optimized for instance to obtain the best performance until 20 GHz. At higher frequencies, one can notice the appearance of parasitic modes and degradation of the circuit performance. More work is needed to extend the useful frequency band above this limitation. Yet the SP4T potentially reach insertion loss better than 0.9 dB with return loss level better than of 20 dB from DC up to 20 GHz. In addition, the isolation level between all accesses is always better than 20 dB in the same frequency band.

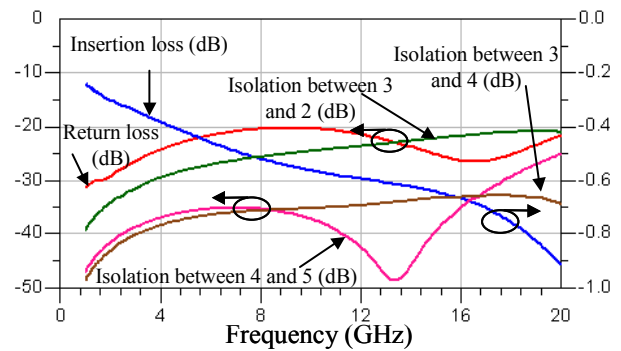

**Fig. 11: Simulated RF performance of the SP4T topology while transmitting from input 1 to output 3.**

## 5. CONCLUSION

In this paper is presented an original packaging approach for the development of RF MEMS switching networks on alumina substrate. This approach has the potentiality to be a low cost solution since it stays at the wafer level and it is compatible with the existing alumina processing technology. The developed packaged MEMS functions seem to present very interesting performances on a wide frequency band. We are currently working on the different device fabrication to validate the simulation results. Latest results will be presented during the conference presentation.